\documentclass[twocolumn,showpacs,amsmath,amssymb]{revtex4}
\input{epsf}

\usepackage{graphicx}
\usepackage{array}
\usepackage{bigstrut}
\usepackage{longtable}
\usepackage{rotating,booktabs}
\usepackage{booktabs,threeparttable}
\usepackage{bm}
\usepackage{lipsum}

\begin{document}

\title{Contributions of negative-energy states to the E2-M1 polarizability difference of the Sr clock}

\author{Fang-Fei Wu,  Ting-Yun Shi, and Li-Yan Tang$^{\dag}$~\footnotetext{\dag Email Address: lytang@apm.ac.cn}}

\affiliation {State Key Laboratory of Magnetic Resonance and Atomic and Molecular Physics, Wuhan Institute of Physics and
Mathematics, Innovation Academy for Precision Measurement Science and Technology, Chinese Academy of Sciences, Wuhan 430071, People's Republic of China}

\date{\today}

\begin{abstract}
With improvement of high-precision optical clock, the higher-order multipolar interaction between atoms and light needs quantitative evaluation. However for the Sr clock, the dynamic E2-M1 polarizability difference at the magic wavelength has contradictions among available results, especially the strongly incompatible sign problem exists between theory and all experiments, which poses new challenges to theory. We investigate contributions of negative-energy states to the E2 and M1 polarizabilities. We find that for the M1 polarizability, the contribution from negative-energy states is dominant. Our result for E2-M1 polarizability difference is $-$7.74(3.92)$\times$10$^{-5}$ a.u., which has the same sign as all the experimental values. Present work has solved the inconsistency problem of sign for E2-M1 polarizability difference in the Sr clock, and the importance of negative-energy-states contribution can be extended directly into the evaluation of multipolar optical frequency shift for other clocks.
\end{abstract}

\maketitle

\section{Introduction}
High-precision optical clock has extensive and important applications, such as redefine the unit of time~\cite{bregolin17a,yamanaka15a,ludlow15a}, establish quantum metrology~\cite{madjarov19a,yin22a}, test variations of the fundamental constants~\cite{godun14a,huntemann14a,safronova18a}, probe dark matter and dark energy~\cite{arvanitaki15a,roberts17a}, and search for new physics~\cite{bars17a,shaniv18a,kolkowitz16a}. Strontium as a typical representative of optical lattice clocks, where atoms are trapped in a magic wavelength optical lattice~\cite{takamoto05a,ludlow06a},
the leading order of Stark shift that related to the dynamic electric dipole (E1) polarizability at the magic wavelength can be eliminated, but the multipolar Stark shifts that related to the dynamic electric quadrupole (E2) and magnetic dipole (M1) polarizabilities can not be cancelled. At present, the systematical uncertainty of the Sr clock has entered into $10^{-18}$ level of precision~\cite{bloom14,nicholson15a,ushijima15a,bothwell19a,oelker19a,lu22a}. Aiming to develop and realize a new generation of higher-precision optical clocks with uncertainty and stability beyond $10^{-18}$, the multipolar interaction between light and atoms related to the E2 and M1 polarizabilities needs to be quantitatively evaluated~\cite{ovsiannikov13a,katori15a,porsev18a,ushijima18a,westergaard11a}.

However, for the Sr clock, there is strongly incompatible sign problem that exists for the E2-M1 polarizability difference at the magic wavelength of 813.4280(5) nm~\cite{ye09a} between theory~\cite{porsev18a, wu2019a,ovsiannikov13a,katori15a} and  experiment~\cite{ushijima18a,dorscher22a,kim22a}, which limits the improvement of precision for the Sr optical clock. At present, the results of two different theoretical methods are consistent with each other. One is from the ab-initio calculations of Porsev {\em et al.}~\cite{porsev18a}, they report a  value of $2.80(36)\times 10^{-5}$ a.u. by using the configuration interaction combined linearized coupled-cluster (CI+all-order) method. The other result of $2.68(94)\times 10^{-5}$ a.u.~\cite{wu2019a}, which is obtained from the combined method of Dirac-Fock plus core polarization (DFCP) and relativistic configuration interaction (RCI) approaches, agrees well with the value of Porsev {\em et al.} But both of theoretical results have opposite sign to the measured value of $-0.962(40)$ mHz of RIKEN group~\cite{ushijima18a}. 

Recently, two newest experimental results for E2-M1 polarizability difference are reported. One value of $-987^{+174}_{-223}$ $\rm{\mu Hz}$ is measured by PTB group~\cite{dorscher22a}, and the other value of $-1.24(5)$ mHz is reported by JILA group~\cite{kim22a}. Both of experiments have same negative sign with the measurement of RIKEN group. This further confirms that the incompatibility of sign between theory and experiment is still pending, which poses a new challenge to theory. Therefore, new theoretical interpretation is urgently needed for solving the current contradiction of the E2-M1 polarizability in the Sr clock.

From theoretical perspective, it is crucial to keep the completeness of intermediates states when using the sum-over-states method to calculate the multipolar polarizabilities. Negative-energy states as products of Dirac theory and as one part of state completeness, the importance of this part has been emphasized in the calculations of $g$-factor of atoms and ions~\cite{shabaev02a,lindroth93a,glazov04a,wagner13a,agababaev18a,arapoglou19a,cakir20a,wu22a}. However, the contribution of negative-energy states to the multipolar polarizabilities for the optical clocks has never been discussed before.

In present work, we take account of the negative-energy-states contributions to the dynamic multipolar polarizabilities of the Sr clock by using improved DFCP+RCI method. Different from available calculations, all the negative-energy states of the Sr$^+$ ion are included to construct configurations of the Sr atom. In addition, the summation in calculating the multipolar polarizabilities involves all the negative-energy and positive-energy states of the Sr atom. We find that for the M1 polarizability, the negative-energy-states contribution is much larger than that of positive-energy states by several orders of magnitude and can completely change the sign of final result. So present work has eliminated the inconsistency problem of sign for E2-M1 polarizability difference in the Sr clock between theory and experiment.   

\section{Theoretical Method}
The combined DFCP+RCI method is effective in predicting structural properties of multi-electron atoms and ions, which can obtain consistent results with other ab-initio methods. For example, for the E1 polarizability of the Sr, Mg, and Cd clocks, the values of DFCP+RCI method agree with the results of CI+all-order method~\cite{wu2019a,wu2020a,zhou2021a} within 3\%. In present work, an improved DFCP+RCI method has been developed by including all the positive- and negative-energy states of monovalent-electron ion to construct the configurations of divalent-electron atom. The detailed implementation process to obtain the energies and wavefunctions of the Sr atom is as follows:

Firstly, we need to solve the Dirac-Fock (DF) equation for the frozen Sr$^{2+}$ core to obtain the real core-orbital
wavefunctions $\psi(\bm{r})$, which can be used to construct the DF potential $V_{DF}(\bm{r})$ between a valence electron and nucleus.

Secondly, we need to solve the DFCP equation to obtain the monovalent-electron wavefunctions $\phi(\bm{r})$ of the Sr$^+$ ion, 
\begin{equation}
	h_{\rm DFCP}(\bm{r})\phi(\bm{r})=\varepsilon\phi(\bm{r})
	\,,\label{e3}
\end{equation}
and $h_{\rm DFCP}(\bm{r})$ represents the DFCP Hamiltonian,
\begin{equation}
	h_{\rm DFCP}(\bm{r})=c{{\bm{\alpha}}}\cdot{\mathbf p}+(\beta-1)c^{2}+V_{N}(\bm{r})+V_{DF}(\bm{r})+V_{\rm 1}(\bm{r})
	\,,	\label{e4}
\end{equation}
where $\bm{\alpha}$ and $\beta$ are the $4\times 4$ Dirac matrices, $\mathbf p$ is the momentum operator for the valence electron, $V_{N}(\bm{r})$ is the Coulomb potential between a valence electron and nucleus. $V_{\rm 1}(\bm{r})$ is the one-body core-polarization potential~\cite{mitroy88c}, which is kept the same as Ref.~\cite{wu2019a}. In this step, it is specially to point out that we keep all the wavefunction $\phi(\bm{r})$ of positive- and negative-energy states for constructing the configuration wavefunctions $|\Phi_{I}(\sigma{\pi} JM)\rangle$ in the following step. 

Thirdly, we perform the configuration interaction calculation for the divalent-electron Sr atom,
\begin{widetext}
\begin{eqnarray}
	\bigg[\sum_i^{2}h_{\rm DFCP}(\bm{r}_i)+\frac{1}{\bm{r}_{12}}+V_{2}(\bm{r}_{12})\bigg]|\Psi({\pi}JM)\rangle=E|\Psi({\pi}JM)\rangle
	\,,\label{e7}
\end{eqnarray}
\end{widetext}
where $V_{2}(\bm{r}_{ij})$ is two-body core-polarization interaction~\cite{mitroy10a,mitroy03f,wu2019a}. The wave function $|\Psi({\pi}JM)\rangle$ with parity $\pi$, angular momentum $J$, and magnetic quantum number $M$ of the system is expanded as a linear combination of the configuration-state wave functions,
\begin{eqnarray}
	|\Psi({\pi}JM)\rangle=\sum_{I}{C_{I}|\Phi_{I}(\sigma{\pi} JM)\rangle}
	\,,\label{e10}
\end{eqnarray}
where $C_{I}$ and $\sigma$ are, respectively, the expansion coefficients and the additional quantum number that define each configuration state uniquely. In this step, we can obtain all the positive- and negative-energy states of the Sr atom. 

When an atom exposed under a linear polarized laser field with the laser frequency $\omega$, the general expressions of dynamic M1 and E2 polarizabilities for the initial state $|0\rangle\equiv|n_0,J_0=0\rangle$ (where $n_0$ represents all other quantum numbers) are written as~\cite{porsev04a}
\begin{eqnarray}
\alpha^{M1}(\omega)&=&\frac{2}{3}\sum_n\frac{\Delta E_{n0}|\langle 0\|T_1^{(0)}\|nJ_n\rangle|^2}{\Delta E_{n0}^2-\omega^2}
\,,\label{e1} \\
\alpha^{E2}(\omega)&=&\frac{1}{30}(\alpha\omega)^{2}\sum_n\frac{\Delta E_{n0}|\langle 0\|T_2^{(1)}\|nJ_n\rangle|^2}{\Delta E_{n0}^2-\omega^2}
\,,\label{e2}
\end{eqnarray}
where $\alpha$ is the fine structure constant, $T^{(\lambda)}_\ell$ is the $2^\ell$-pole transition operator, $\lambda=0$ and $\lambda=1$ represent the magnetic and electric transition operators, respectively. $\Delta E_{n0}$ represents the transition energy between the initial state $|0\rangle$ and the intermediate state $|nJ_n\rangle$. 
The reduced matrix elements $\langle 0\|T_1^{(0)}\|nJ_n\rangle$ and $\langle 0\|T_2^{(1)}\|nJ_n\rangle$ can be expressed by the reduced matrix elements $\langle i\|t_1^{(0)}\|k\rangle$ and $\langle i\|t_2^{(1)}\|k\rangle$ of monovalent-electron system~\cite{johnson06a}, 
\begin{eqnarray}
	\langle i\|t^{(0)}_1\|j\rangle&=&\frac{\kappa_i+\kappa_j}{2}\langle-\kappa_i\|C^1\|\kappa_j\rangle \nonumber \\
	&&\int r[P_i(r)Q_j(r)+Q_i(r)P_j(r)]dr
	\,,\label{e4}
\end{eqnarray}
\begin{eqnarray}
	\langle i\|t^{(1)}_2\|j\rangle&=&\langle\kappa_i\|C^2\|\kappa_j\rangle \nonumber \\
	&&\int r^2[P_i(r)P_j(r)+Q_i(r)Q_j(r)]dr
	\,,\label{e5}
\end{eqnarray}
where $P_i(r)$ and $Q_i(r)$ are the large and small components of wavefunctions for monovalent-electron system. It is worth noting that the radial integrations of electric and magnetic reduced matrix elements are different. The magnetic part in Eq.~(\ref{e4}) involves the cross product of large and small components of wavefunctions. In addition, the summation of the M1 and E2 polarizabilities in Eqs.~(\ref{e1}) and~(\ref{e2}) involves all the intermediate states, including the negative-energy states. 

In present work, we have tested the convergence of multipolar polarizabilities as the number of B-spline basis sets $N$ and partial-wave $\ell$ increased. We find that our results remain unchanged with at less 4 significant digits as $N$ and $\ell$ increased. So in the following section, we choose to list the results under the maximum basis set. The maximum number of B-spline basis in our calculation is 40, the maximum number of partial-wave is 5, and the total number of configuration has reached 128781.

\section{Results and Discussions}
Using the improved DFCP+RCI method with negative-energy states included, we have performed calculations of energies, reduced matrix elements,  and multipolar polarizabilities for the Sr clock. We find that with inclusion of the negative-energy states, the energy correction for low-lying states is less than 3 ppm. And the negative-energy states have little effect on the E1 polarizability, which can not be reflected under current theoretical accuracy.

For the multipolar polarizabilities of the clock states at the 813.4280(5) nm magic wavelength that we are interested in, Tables~\ref{t1} and \ref{t2} list the itemized contributions to the E2 and M1 polarizabilities, respectively. It is seen that for the E2 polarizability, the main contribution to the ground-state $5s^2\,^1S_0$ comes from the positive-energy states of $5s4d\,^1D_2$ and $5s5d\,^1D_2$, both of them contribute about 75\% and 13\% to the total E2 polarizability, respectively. For the excited-state $5s5p\,^3P_0^o$, the main contribution comes from the $5d5p\,^3F_2^o$, $5s6p\,^3P_2^o$ and $5s4f\,^3F_2^o$ states, these three items together contribute about 60\% to the total E2 polarizability. Especially, for both of the $5s^2\,^1S_0$ and $5s5p\,^3P_0^o$ clock states, the contribution of negative-energy states is less than $10^{-14}$, which can be almost ignored.

\begin{table}[!htbp]
\caption{\label{t1} Itemized contributions (Contr.) to the dynamic E2 polarizability (in a.u.) for the $5s^2\,^1S_0$ and $5s5p\,^3P_0^o$ clock states at the 813.4280(5) nm magic wavelength. Tail represents the contribution from other positive-energy states, $\alpha^{E2+}$ and $\alpha^{E2-}$ represent the total contribution from positive-energy and negative-energy states, respectively. The numbers in the square brackets denote powers of ten.}
\begin{ruledtabular}
\begin{tabular}{llll}
\multicolumn{2}{c}{$5s^2\,^1S_0$}&\multicolumn{2}{c}{$5s5p\,^3P_0^o$}\\
\cline{1-2}\cline{3-4}
\multicolumn{1}{c}{Sub item} &\multicolumn{1}{c}{Contr.}
&\multicolumn{1}{c}{Sub item} &\multicolumn{1}{c}{Contr.}\\ \hline
$5s4d\,^3D_2$   &1.258[-7]     &$5s5p\,^3P_2^o$   &$-$2.805[-6]  \\
$5s4d\,^1D_2$   &6.965[-5]     &$5d5p\,^3F_2^o$   &3.095[-5]    \\
$5s5d\,^1D_2$   &1.224[-5]     &$5d5p\,^1D_2^o$   &3.149[-6]  \\
$5s5d\,^3D_2$   &1.106[-8]     &$5s6p\,^3P_2^o$   &1.741[-5]  \\
$5p^2\,^3P_2$   &5.966[-8]     &$4d5p\,^3D_2^o$   &3.603[-6]  \\
$5d^2\,^1D_2$   &3.887[-8]     &$5d5p\,^3P_2^o$   &2.139[-6]  \\
$5s6d\,^3D_2$   &4.981[-10]    &$5s4f\,^3F_2^o$   &2.644[-5]  \\
$5s6d\,^1D_2$   &1.226[-7]     &$5s7p\,^3P_2^o$   &2.601[-6]  \\
$5s7d\,^1D_2$   &2.600[-6]     &$5s5f\,^3F_2^o$   &8.768[-6]  \\
Tail            &7.950[-6]     &Tail              &3.214[-5]        \\
$\alpha^{E2+}$  &9.28[-5]      &$\alpha^{E2+}$    &12.44[-5]     \\
$\alpha^{E2-}$  &$-$8.64[-16]  &$\alpha^{E2-}$    &$-$1.10[-15]   \\
Total           &9.28[-5]      &Total             &12.44[-5]      \\
\end{tabular}
\end{ruledtabular}
\end{table}
\begin{table}[!htbp]
\caption{\label{t2} Itemized contributions (Contr.) to the dynamic M1 polarizability (in a.u.) for the $5s^2\,^1S_0$ and $5s5p\,^3P_0^o$ clock states at the 813.4280(5) nm magic wavelength. Tail represents the contribution from other positive-energy states, $\alpha^{M1+}$ and $\alpha^{M1-}$ represent the total contribution from positive-energy and negative-energy states, respectively. The numbers in the square brackets denote powers of ten.}
\begin{ruledtabular}
\begin{tabular}{llll}
\multicolumn{2}{c}{$5s^2\,^1S_0$}&\multicolumn{2}{c}{$5s5p\,^3P_0^o$}\\
\cline{1-2}\cline{3-4}
\multicolumn{1}{c}{Sub item} &\multicolumn{1}{c}{Contr.}
&\multicolumn{1}{c}{Sub item} &\multicolumn{1}{c}{Contr.}\\ \hline
$5s4d\,^3D_1$    & 1.483[-15]      &$5s5p\,^3P_1^o$ &$-$4.811[-6] \\
$5s6s\,^3S_1$    & 4.098[-13]      &$5s5p\,^1P_1^o$ &$-$2.702[-7] \\
$5s5d\,^3D_1$    & 1.273[-12]      &$5s6p\,^3P_1^o$ &7.336[-10]    \\
$5p^2\,^3P_1$    & 1.539[-9]       &$5s6p\,^1P_1^o$ &1.766[-8]       \\
Tail             & 5.81[-10]       &Tail            &1.35[-8]     \\
$\alpha^{M1+}$   & 2.17[-9]        &$\alpha^{M1+}$  &$-$5.05[-6]   \\
$\alpha^{M1-}$   & $-$3.84[-4]     &$\alpha^{M1-}$  &$-$4.88[-4]   \\
Total            & $-$3.84[-4]     &Total           &$-$4.93[-4]   \\
\end{tabular}
\end{ruledtabular}
\end{table}

However, different from the E2 polarizability, the influence of negative-energy states on the dynamic M1 polarizability is obvious and dominant, which can be seen clearly from Table~\ref{t2}. For the $5s^2\,^1S_0$ state, if the negative-energy states are not taken into account, the largest contribution comes from the $5p^2\,^3P_1$ state. After considering the negative-energy states, the final M1 polarizability at the 813.4280(5) nm magic wavelength is changed from $2.17\times 10^{-9}$ a.u. to $-3.84\times 10^{-4}$ a.u., the sign of which is changed completely. This dues to the contribution of negative-energy states is five orders of magnitude larger than that of the positive-energy states, and the sign of contribution is opposite. Similarly, for the $5s5p\,^3P_0^o$ state, the contribution of negative-energy states is two orders of magnitude larger than that of the positive-energy states, accounting for 99\% of the final M1 polarizability. 

In order to further explore the reasons of large and dominant negative-energy-states contribution, we have analyzed the itemized contributions of negative-energy states. We find that different from the positive-energy-states contribution, the negative-energy-state contribution is not the main contribution of a few intermediate states, but is a cumulative effect of thousands of negative-energy states with energy in the range of $-37558(1)$ ($2mc^2\approx37558$) a.u. Although all these negative-energy states of $-37558(1)$ a.u. are far away from the initial state, the radial wavefunction $Q_j(r)$ of these states have large overlap with the $P_i(r)$ part of the initial state wavefunction, which results in the large $P_i(r)Q_j(r)$ product in Eq.~(\ref{e4}). In other words, it is a series of large M1 transition matrix elements between the negative-energy states and the initial state that lead to the dominant contribution of negative-energy states to the M1 polarizability.

\begin{table}[ht]
\caption{\label{t3} Summarized results of dynamic E2 and M1 polarizabilities (in a.u.) for the $5s^2\,^1S_0$ and $5s5p\,^3P_0^o$ clock states at the 813.4280(5) nm magic wavelength. $\Delta\alpha^{E2}(\omega)$ and $\Delta\alpha^{M1}(\omega)$ represent the difference for the clock states of the dynamic E2 and M1 polarizabilities, respectively. And $\Delta\alpha^{QM}(\omega)=\Delta\alpha^{M1}(\omega)+\Delta\alpha^{E2}(\omega)$. The numbers in parentheses are computational uncertainties. The numbers in the square brackets denote powers of ten.}
\begin{ruledtabular}
\begin{tabular}{lrrr}
\multicolumn{1}{l}{Polarizability}&\multicolumn{1}{c}{Present}&\multicolumn{1}{c}{Ref.~\cite{wu2019a}}
&\multicolumn{1}{c}{Ref.~\cite{porsev18a}} \\ \hline
  $\alpha^{E2}_{\,^1S_0}(\omega)$   &  9.28(57)[-5]       &9.26(56)[-5]     &8.87(26)[-5]  \\
\specialrule{0em}{1pt}{1pt}
  $\alpha^{E2}_{\,^3P_0^o}(\omega)$ &  12.44(76)[-5]      &12.44(76)[-5]    &12.2(25)[-5]  \\
\specialrule{0em}{1pt}{1pt}
  $\Delta\alpha^{E2}(\omega)$       &  3.16(95)[-5]       &3.18(94)[-5]     &3.31(36)[-5]  \\ \hline
\specialrule{0em}{1pt}{1pt}
  $\alpha^{M1}_{\,^1S_0}(\omega)$   & $-$3.84(24)[-4]     &2.12(13)[-9]     &2.37[-9]      \\
\specialrule{0em}{1pt}{1pt}
  $\alpha^{M1}_{\,^3P_0^o}(\omega)$ & $-$4.93(30)[-4]     &$-$5.05(31)[-6]  &$-$5.08[-6]   \\
\specialrule{0em}{1pt}{1pt}
  $\Delta\alpha^{M1}(\omega)$       & $-$1.09(38)[-4]     &$-$5.05(31)[-6]  &$-$5.08[-6]   \\  \hline
\specialrule{0em}{1pt}{1pt}
  $\Delta\alpha^{QM}(\omega)$       & $-$7.74(3.92)[-5]   &2.68(94)[-5]     &2.80(36)[-5]  \\
\end{tabular}
\end{ruledtabular}
\end{table}
\begin{figure}
\includegraphics[width=0.52\textwidth]{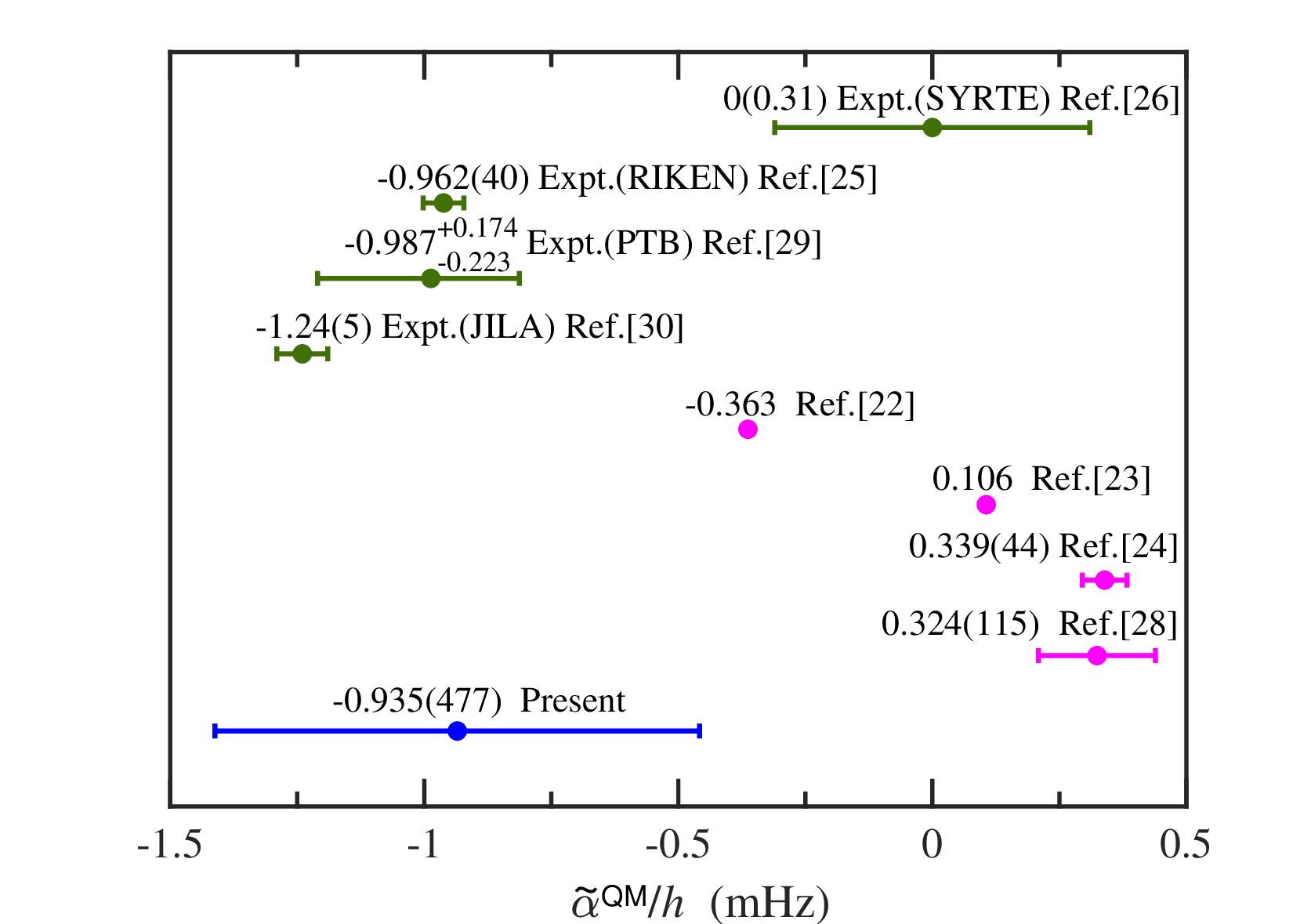}
\caption{\label{f1}(Color online) Comparison of the $\tilde{\alpha}^{QM}/h$ (in mHz). The green line represents experimental results. The blue line represents our present value, and the magenta line denotes other theoretical results.}
\end{figure}

Further, since the results of DFCP+RCI method are consistent with that of ab-initio calculations within 3\%~\cite{wu2019a,wu2020a}, we can introduce $\pm$3\% fluctuation into all the reduced matrix elements to conservatively evaluate the uncertainty of present E2 and M1 polarizabilities. The final values are summarized in Table~\ref{t3}, and a detailed comparison is also given in this table. It can be clearly seen that for the E2 polarizability, present values including the negative-energy-states contribution are in good agreement with Refs.~\cite{porsev18a,wu2019a}, which only include the positive-energy states. This confirms again that the negative-energy-states contribution to the electric polarizability can be neglected.

In addition, the obvious difference between present work and other calculations in Table~\ref{t3} is the M1 polarizability. For the $5s^2\,^1S_0$ state, present value of $-3.84(24)\times 10^{-4}$ a.u. has opposite sign to the values of Refs.~\cite{porsev18a,wu2019a}. And the absolute value of $-3.84(24)\times 10^{-4}$ a.u. is five orders of magnitude larger than the values of Refs.~\cite{porsev18a,wu2019a}. For the $5s5p\,^3P_0^o$ state, present value of $-4.93(30)\times 10^{-4}$ a.u. is two orders of magnitude larger than other values in Refs.~\cite{porsev18a,wu2019a}. 

When adding $\Delta\alpha^{E2}(\omega)$ and $\Delta\alpha^{M1}(\omega)$ together, we can get the final E2-M1 polarizability difference $\Delta\alpha^{QM}(\omega)=-7.74(3.92)\times10^{-5}$ a.u., which includes the negative-energy-states contribution of $-1.04(38)\times 10^{-4}$ a.u. Compared with our previous value of $2.68(94)\times 10^{-5}$ a.u.~\cite{wu2019a}, the large uncertainty in present work dues to the dominant differential M1 polarizability $\Delta\alpha^{M1}(\omega)$. Since the absolute value of $-1.09(38)\times 10^{-4}$ a.u. is an order of magnitude larger than the differential E2 polarizability $\Delta\alpha^{E2}(\omega)=3.16(95)\times 10^{-5}$ a.u., the addition of two terms causes the cancellation of significant figures. This is completely different from other calculations of Refs.~\cite{porsev18a,wu2019a}, where $\Delta\alpha^{M1}(\omega)$ is an order of magnitude less than $\Delta\alpha^{E2}(\omega)$. For the larger uncertainty in our present value of $-7.74(3.92)\times10^{-5}$ a.u., there is limited room to improve the accuracy for our DFCP+RCI method at present. Therefore, to further reduce the theoretical uncertainty in the future, it is necessary to develop high-accuracy theoretical methods for calculations of multi-electron atomic structure, such as the CI+all-order method.

To compare with experiments directly, we convert all the theoretical values of E2-M1 polarizability difference from atomic units (a. u.) to the unit of Hz. It is can be seen from Fig.~\ref{f1}. Where $\tilde{\alpha}^{QM}=\Delta \alpha^{QM}(\omega)E_R/\alpha^{E1}(\omega)$, $\alpha^{E1}(\omega)=287(17)$ a.u. is present dynamic E1 polarizability at 813 nm magic wavelength of clock states, and $E_R$ is the lattice photon recoil energy~\cite{ushijima18a}. It is clearly seen that our present value of $-0.935(477)$ mHz, which includes the negative-energy-states contribution, agrees well with the three measured results of $-0.962(40)$~\cite{ushijima18a}, $-0.987^{+0.174}_{-0.223}$~\cite{dorscher22a} and $-1.24(5)$ mHz~\cite{kim22a}. This illustrates that the negative-energy states are crucial to the calculation of multipolar polarizabilities.

In addition, from Fig.~\ref{f1}, it is also seen that there discrepancy exists between the recent measurement of JILA~\cite{kim22a} and previous measurement of RIKEN~\cite{ushijima18a}. If adding present negative-energy-states contribution $-1.04(38)\times 10^{-4}$ a.u. into the CI+all-order value of $2.80(36)\times10^{-5}$ a.u.~\cite{porsev18a}, and considering the uncertainty in CI+all-order method is about a factor of 1/3 of our DFCP+RCI method, then we can get an estimated value of $-7.60(1.50)\times 10^{-5}$ a.u. (equals to $-0.918(189)$ mHz), which is expected to judge on the current experimental results after taking into account of the negative-energy-states contribution. Therefore, development of high-accuracy theoretical methods with negative-energy states included is urgently needed to solve the existing discrepancy among different measurements.

\section{Conclusions}
Focusing on the obvious contradiction in sign for the E2-M1 polarizability difference between existing theory and experiment in the Sr clock, we develop the combined DFCP+RCI method with inclusion of negative-energy states, and apply it into calculations of dynamic M1 and E2 polarizabilities for the Sr clock. Our result of E2-M1 polarizability difference is $-$7.74(3.92)$\times$10$^{-5}$ a.u., which has the same sign with all the measured values. Our work has solved the sign inconsistency for the E2-M1 polarizability difference in the Sr clock. In the future, developing high-accuracy theoretical method with the negative-energy states included is expected to solve the discrepancy among different experiments. In addition, our work has revealed the importance of negative-energy states that lack in all previous calculations of optical clocks, which can be extended into investigations of multipolar interaction between atoms and light in the field of precision measurement physics.

\begin{acknowledgments}
We thank Yong-Hui Zhang for helpful discussions on the negative-energy states, and thank J. Chen, K.-L. Gao, and Z.-C. Yan for reading our paper. This work was supported by the National Natural Science Foundation of China under Grant Nos. 12174402 and 12004124, and by the Nature Science Foundation of Hubei Province Nos.2019CFA058 and 2022CFA013.
\end{acknowledgments}


\end{document}